# Nanoporous Carbon Nitride: A High Efficient Filter for Seawater Desalination

Weifeng Li, [1*‖] Yanmei Yang, [1‖] Hongcai Zhou, [2] Xiaoming Zhang [2] and Mingwen Zhao [2*]

[1] School for Radiological and Interdisciplinary Sciences (RAD-X) and Collaborative Innovation Center of Radiation Medicine of Jiangsu Higher Education Institutions, Soochow University, Suzhou, China, 215123;

[2] School of Physics and State Key Laboratory of Crystal Materials, Shandong University, Jinan, China, 250100

**Abstract:** The low efficiency of commercially-used reverse osmosis (RO) membranes has been the main obstacle in seawater desalination application. Here, we report the auspicious performance, through molecular dynamics simulations, of a seawater desalination filter based on the recently-synthesized graphene-like carbon nitride (g-$C_2$N) [***Nat. Commun.,*** 2015, 6, 6486]. Taking advantage of the inherent nanopores and excellent mechanical properties of g-$C_2$N filter, highly efficient seawater desalination can be achieved by modulating the nanopores under tensile strain. The water permeability can be improved by two orders of magnitude compared to RO membranes, which offers a promising approach to the global water shortage solution.

Corresponding authors: W. F. Li: wfli@suda.edu.cn; M. W. Zhao: zmw@sdu.edu.cn

‖ These authors contributed equally.

Because of the surging growth of population, rapid urban development and industrialization progress, the augmentation in available freshwater resources is dwindling in many water-stressed countries of the world.[1] Seawater represents a plentiful resource to compensate the stress of potable water supply through desalination. Since the early 1960s, the most prevalent desalination method has been reverse osmosis (RO) technique.[2] In this strategy, a semi-permeable membrane is placed at the interface between seawater and pure water. Pressure applied at the seawater side forces water flow towards the pure water side, while ions are blocked. The RO method is prohibitively energy intensive at large water volumes, due to the low water production rate of only ~0.1 L/cm$^2$/day/MPa for commercially-used RO membranes.[1,3] The sustainability of desalinated water resources, therefore, likely depends on the development of new desalination technologies.

An alternate approach is the graphene filters which were first proposed from computer simulations[3-6] and eventually realized in recent experiments.[7,8] The ultrathin thickness (only one atomic layer) of graphene lowers energy consumption because water flux scales inversely with the filter thickness. This approach may also work for other two-dimensional (2D) materials, such as hexagonal boron nitride,[9] silicene[10,11] and phosphorene.[12-15] However, these 2D materials face the same limitation as graphene: they do not have suitable nanopores inherently for water flow and thus post-treatment is essential. Control of the shape and size of the nanopores which are decisive for water transparency and ion selectivity remains challenging in post-treatment experiments. Approaches to tune the nanopores continuously to meet the requirement of highly-efficient seawater desalination are lacking.

A promising starting point to this goal will be a 2D material with natural porous framework. We noticed that the recently-synthesized graphene-like carbon nitride[16] (g-C$_2$N, as shown in Fig. 1) has natural nanopores of 5.5 Å (measured from the N-N distance), close to the diameter of a water molecule (~4 Å). The stability the g-C$_2$N membrane at high temperatures has also been evidenced[16]. Moreover, the covalent bond (C-C and C-N) network of the g-C$_2$N would lead to excellent mechanical properties comparable to graphene, which can sustain large tensile strain. It is apparent that the g-$C_2N$ membrane is ready for the desalination test, where molecular dynamics simulation could play the exploratory role.

In this contribution, we performed theoretical simulations on the seawater desalination efficiency of the g-$C_2N$ membrane under tensile strain. It is found that modulated by tensile strain the transmission of water molecules through the nanopores can be tuned continuously from "close" to "open" states, leading to a water permeability of two orders of magnitude higher than that of the RO membranes, while ion transmission is blocked. We attributed it to the strain-modulated steric hindrance effect on the transmission of water molecules and ions through nano-confined spaces. In view of the abundant porous structures of graphene-like carbon nitrides[17-22], these results offer not only an efficient strategy to modulate water transmission behaviors in confined spaces but also a promising approach to the global water shortage solution.

As illustrated in Fig. 1a, the simulation model was composed of a seawater region and a pure water region, separated by a g-$C_2N$ filter. A graphene sheet mimicking the piston was placed on the top of the seawater and force in the z direction was loaded on it to push seawater flow towards the pure water region. The g-$C_2N$ filter (Fig. 1b) containing 30 nanopores has a dimension of roughly 4.16×4.33 $nm^2$ at the equilibrium state (details see the computational section).

We first tested the water transparency through the g-$C_2N$ filter in response to biaxial tensile strain ($\tau$). The cumulative numbers of water molecules transferred through the g-$C_2N$ filter under a piston pressure of 100 MPa are summarized in Fig. 2a. Under weakly strained ($\tau \leq 1\%$) conditions, no event of water passage is observed during the entire trajectories, corresponding to the "closed" state of g-$C_2N$ filter. The steric hindrance of the nanopores blocks the transmission of water molecules at this stage. The transition point from "closed" to "open" appears at a strain level near 2%, as a small amount of water does pass through the filter. As the strain is further increased, the g-$C_2N$ filter becomes more transparent. For $\tau \geq 6\%$, each curve begins with a linear regime and eventually reaches a saturation point (around 6500 water molecules) where the entire reservoir of water molecules on the sea water region is depleted. More importantly, for all the simulations under various strain strength, the water flow (the slope of curves in Fig. 2a) is constant in time, confirming that well-converged statistics were obtained for all the simulations.

The water flow through a 12%-stretched g-$C_2N$ filter as a functional of piston pressure is summarized in Fig. 2b. It is clear that the water flow is proportional to the strength of applied pressure. This allows us to evaluate the water permeability of g-$C_2N$ filter by extrapolating the dynamic quantities derived here down to the operating condition that is more typical of reverse osmosis plants (usually several MPa). We expressed the water permeability in liters of output per square centimeter of the filter per day and per unit of applied pressure. As shown in Fig. 2b, it varies from zero (for the unstrained g-$C_2N$) to 35.1 L/$cm^2$/day/MPa (for the extremely strained case 12%). The high performance benefits from the packed permeable pores in g-$C_2N$ filter (density of $1.6\times10^{14}$ $cm^{-2}$). For comparison, the highest performance that can be achieved for graphene filter is around 6 L/$cm^2$/day/MPa[7] which is only 1/6 of that of g-$C_2N$ filter. This solidly highlights the significance of the unique porous structure of g-$C_2N$. It is also noteworthy that the performance of commercial RO is only in the order of 0.1 L/$cm^2$/day/MPa.[5, 23] The g-$C_2N$ filter revealed in our simulations, however, has a super high water permeability of two orders of magnitude higher than commercial RO desalination membrane. Particularly, these exists a window (6%-11%) of tensile strain where the water permeability scales almost linearly with strain strength as demonstrated in Fig. 2b. This window allows precise control of the g-$C_2N$ filter which is quite crucial for the design of tunable devices.

The strain-regulated water flux through the nanopores of the g-$C_2N$ filter may have two origins: (1) the confinement effect of the nanopores to water molecules; (2) the charge redistribution at the edges of the nanopores. It has been demonstrated that the edge structure of the nanopores affects the water flow. For example, the graphene filters with hydrogen or hydroxyl functionalization at the pore edges have quite distinct water permeability.[24] The hydroxyl groups can roughly double the water flux, compared to the hydrogen case, due to their hydrophilic character. Based on the knowledge that stretching of chemical bonds may result in charge redistribution between two adjacent elements (typically around 0.1 |e|), we change the charges of the nitrogen atoms at the pore edges of g-$C_2$N from 0.4 to 0.6 |e| in our simulations. The water permeabilities of the g-$C_2$N filter under tensile strain of 6% and 11% are summarized in Fig. 3. It is clear that the filter with less-charged nanopores has larger water permeability. On the contrary, accumulation of electrons at the nanopores will effectively decrease the water permeability. We attribute this phenomenon to the attraction of negatively-charged nitrogen atoms at the edges to water molecules which decreases the tunnels' effetive cross-section for water passage. Nitrogen atoms with larger charges are capable to attract water in a firm pattern, leaving a narrower tunnel for water flow. However, the change of water permeability in response to the charges of nitrogen is only 1.0 L/cm$^2$/day/MPa per 0.1|e|, much lower than the strain effect. Therefore, it is safe to propose that the regulation of the water flux by strain is mainly determined by the changes of steric hindrance upon modulations of the nanopore size.

Beside high water permeability, a desalination filter should effectively hinder the passage of salt ions. The desalination efficiency is determined by the trade-off between water permeability and salt ion selectivity. However, the nanopores generated in pre-treatment of graphene always have a wide range of size, which limits the efficiency of graphene filters. It is exciting that no events of ion passage have been observed through the g-$C_2$N filter throughout the simulations with serious of strain and pressure, even for the extreme case of 12%-strained filter. This indicates the superior salt ion selectivity of g-$C_2$N filter, which is quite crucial for desalination performance improvement.

The superior desalination performance of g-$C_2$N filter can be explained by the potential of mean force (PMF) analysis through umbrella sampling. The PMF curves for $Na^+$, $Cl^-$, and water to across the g-$C_2$N filter were obtained by sampling the force experienced by the salt ions or water molecules when passing through the nanopores. Without losing the validity, we have only calculated the PMF curves for the

12%-stretched g-$C_2N$. As shown in Fig. 2d, the PMF for water molecules is most shallow, with no energy barrier (peak) or well (valley) exceeding 3.32 $k_BT$. Typically, an energy barrier of around 5 $k_BT$ is considered to be low enough for permeation to happen[25-27]. Hence water molecules can pass through the nanopores easily, leading to high water transparency. $Na^+$ is blocked because of the high energy barrier of 12.27 $k_BT$ near the nanopore center. The probability ($K_r$) for certain solution component to overcome an energy barrier ($E_b$) , $K_r \propto exp((-E_b)/k_BT)$, indicates that the water molecules have approximately 8×10³ times higher chance to pass through the g-$C_2N$ nanopores compared to $Na^+$ ions at room temperature. For $Cl^-$ to approach the nanopores from the sea water side, an energy barrier of higher than 35 $k_BT$ was predicted which is almost inaccessible for transmission to happen. Such high barriers for salt ions can be attributed to two characteristics of the nanopores in g-$C_2N$ filter. Firstly, the water-permeable nanopores are relatively small (d ~ 5.5Å), thus the steric exclusion effects of the nanopores disfavor ion translocation. Secondly and most importantly, the pyridine nitrogen atoms at the nanopore edge are negatively charged (in total, by about -2.88|e|) which present strong electronic repulsion to the anionic passage.

The stability of g-$C_2N$ filter under tensile strain is another important factor that affects the desalination efficiency. We calculated the strain energy ($E_s$) with respect to the equilibrium state from first-principles, as shown in Fig. 4a. With the increase of tensile strain, $E_s$ increases monotonously as τ < 20%. But the derivate of strain energy ($dE_s/d\tau$) reaches the maximum value at tensile strain of around 13%, suggesting that the g-$C_2N$ lattice becomes softer at this stage. This is also consistent with the features of phonon spectra. When the tensile strain is smaller than 12%, the phonon spectra are free from imaginary frequency modes, as shown in Fig. 4b, which confirms the stability of the filter at this strain range. As the tensile strain exceeds this critical point, imaginary frequency modes appear as shown in Fig. 4c. Although the covalent bonds are not broken completely at this stage, a small external disturbance may destroy the g-$C_2N$ framework due to the imaginary frequency modes. Both the energy evolution and phonon spectra confirm that the g-C$_2$N filter is mechanically stable under tensile strain lower than 12%, which can be attributed to the strong covalent bond framework that is comparable to graphene. The excellent mechanical properties of the g-C$_2$N lattice fulfill the requirement of filters working at high pressure. It should be mentioned that although the g-C$_2$N lattice can sustain large tensile up to 12%, we needn't to apply such large tensile strain to the g-C$_2$N filter to get high water desalination efficiency. For example, at moderate tensile strain (~7%), the water permeability, ~10 L/cm$^2$/day/MPa, is higher than that of the commercial RO by about two orders.

In summary, our theoretical simulations imply that the recently-synthesized g-C$_2$N membrane can serve as an efficient filter for seawater desalination. The natural nanopores of g-C$_2$N filter can conduct water in a high transparency manner, while salt ions are completely rejected. The water permeability is estimated to be two orders of magnitude higher than the commercially used RO membranes and six times higher than graphene filters. The switch between "open" and "closed" states for water flow can be achieved by applying tensile strain. The accurate regulation of the filter under tensile strain and the precise pressure-responsive behavior make the g-C$_2$N on the horizon to advance the development of desalination membrane design. Our results also support the design and proliferation of tunable devices for filtration and other applications working at nanoscale.

*Computational details*

All the simulations were performed with the GROMACS package.[28] The AMBER03 force field[29] was used in the simulations. The seawater region contains 50 Na$^+$ and Cl$^-$ ions, and 7000 TIP3P water molecules,[30] corresponding to a salt concentration of 27 g/L, slightly lower than the salinity of seawater (~35 g/L). A lower salinity was chosen for the consideration that water flowing towards the pure water region will effectively increase the salinity in seawater side during the simulation. For the g-C$_2$N filter, the atom types of CA and NB were assigned to C and N atoms. The RESP point charges were calculated using the Gaussian 09 code[31] at the HF/6-31G* level, yielding a value of 0.24|e| and -0.48|e| for C and N atoms, respectively. SHAKE constraints[32] were applied to all bonds involving hydrogen atoms. The long-range electrostatic interactions were treated with the Particle Mesh Ewald method,[33,34] and atypical distance cutoff of 12 Å was adopted for the van der Waals (vdW) interactions. The non-bonded interaction pair list was updated every 10 fs. The cross section in the *x-y* plane of the simulation box was fixed to a certain value in order to mimic the strained filter. The box was coupled to a constant at 1.0 atm only along the *z* direction. Canonical sampling was performed through velocity rescaling method at constant temperature 300 K.[35] A movement integration step of 1 fs was used in the simulations. Each system was first equilibrated for 10 ns, followed by 300 ns productive simulation for the data collection.

First-principles calculations were performed using the Vienna Ab initio Simulation Package (VASP)[36-38]. The electron-electron interactions are treated using a generalized gradient approximation (GGA) in the form of Perdew-Burke-Ernzerhof (PBE) for the exchange-correlation functional[39]. The energy cutoff of the plane waves was set to 520 eV with an energy precision of $10^{-8}$ eV. Vacuum space larger than 15 Å was used to avoid the interaction between adjacent images. The Monkhorst-Pack meshes of 9×9×1 were used in sampling the Brillouin zone for the g-$C_2N$ lattice. Tensile strain was applied by fixing the lattice constants to different values. Atomic coordinates were optimized using the conjugate gradient (CG) scheme until the maximum force on each atom was less than 0.01eV/Å. Phonon spectra were calculated using a supercell approach within the PHONON code.[40]

**ACKNOWLEDGMENT**

This work was supported by the National Basic Research Program of China (No. 2012CB932302), the National Natural Science Foundation of China (grant no. 11304214, 21405108, and 21433006), the Priority Academic Program Development of Jiangsu Higher Education Institutions (PAPD).

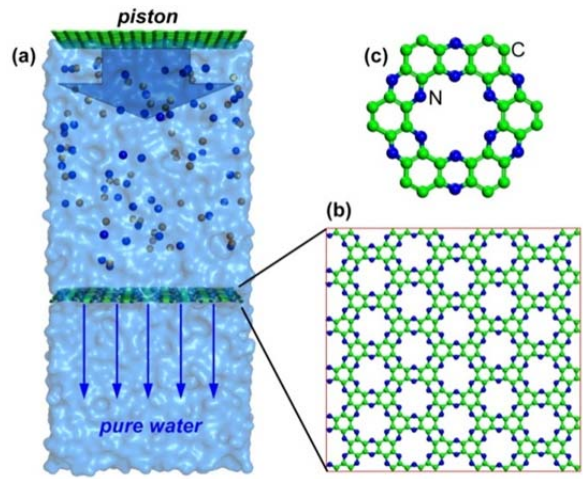

**Figure 1**.(a) Illustration of the simulation model; (b) Top view of the nanoporous g-$C_2$N layer; (c) Local structure of one nanopore in g-$C_2$N.

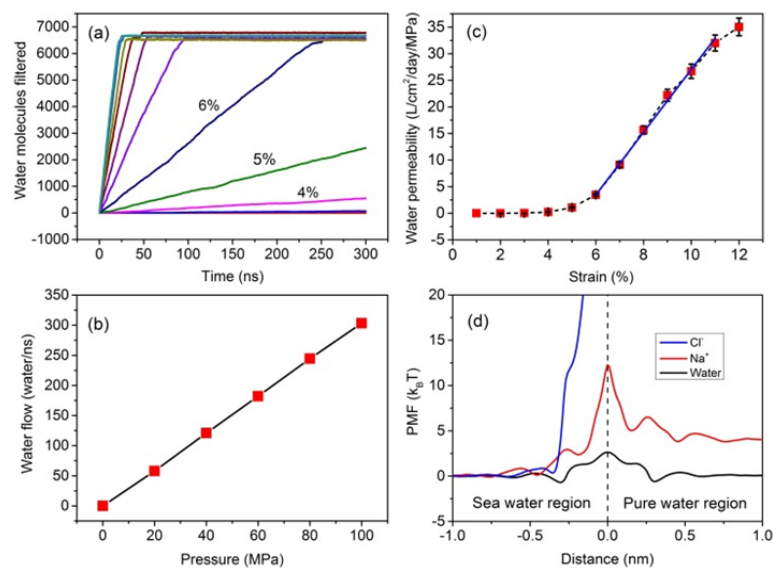

**Figure 2**. (a) Number of water filtered by g-$C_2$N membrane as a function of simulation time under piston pressure of 100 MPa, (b) Water flow at various pressure through the 12%-stretched g-$C_2$N, (c) Water permeability with respect to tensile strain and (d) Potential of mean force (PMF) of water, $Na^+$ and $Cl^-$ passing through the 12%-stretched g-$C_2$N filter from seawater region to pure water region. The filter is placed at 0 nm.

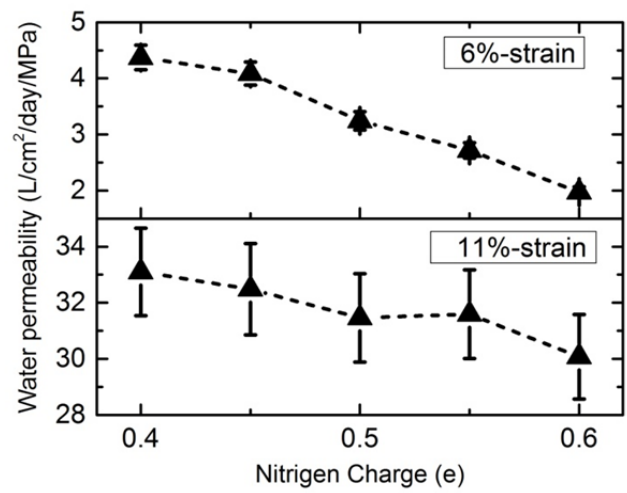

**Figure 3**. Water permeability with respect to the charges of nitrogen in the g-$C_2N$ filter under tensile strain of 6% and 11%.

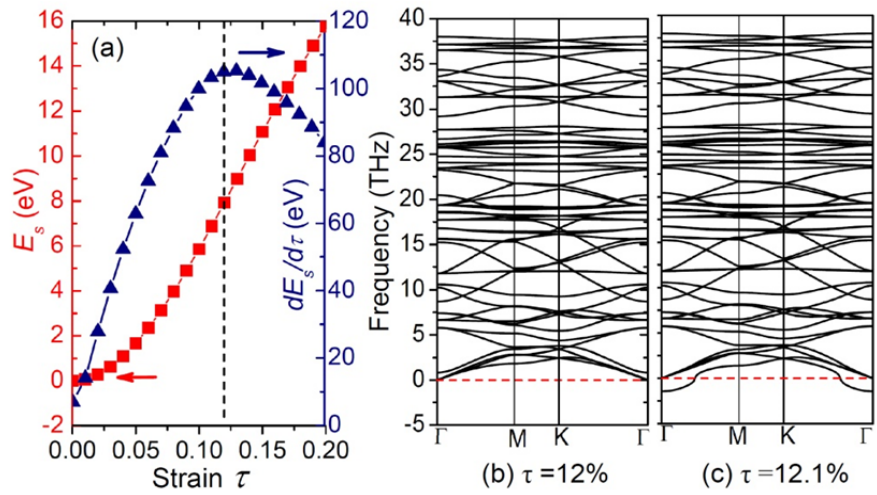

Figure 4. (a) Variation of energy in response to tensile strain. (b)-(c) Phonon spectra of g-$C_2N$ under the biaxial tensile strains of 12% and 12.1%. The negative frequencies correspond to the imaginary frequency modes which are dynamically unstable.